\documentclass[fleqn,usenatbib]{mnras}

\usepackage{newtxtext,newtxmath}

\usepackage[T1]{fontenc}

\DeclareRobustCommand{\VAN}[3]{#2}
\let\VANthebibliography\thebibliography
\def\thebibliography{\DeclareRobustCommand{\VAN}[3]{##3}\VANthebibliography}

\usepackage{graphicx}	
\usepackage{amsmath}	
\usepackage{hyperref}
\usepackage{orcidlink}

\title[Distances to KUV 00311-1938 and S2 0109+22]{Galaxy-group-associated distances to Very High Energy gamma-ray emitting BL Lacs KUV 00311-1938 and S2 0109+22}

\author[K. I. I. Koljonen et al.]{Karri I. I. Koljonen$^{\orcidlink{0000-0002-9677-1533}}$,$^{1,2,3}$\thanks{E-mail: karri.koljonen@ntnu.no}
Elina Lindfors$^{\orcidlink{0000-0002-9155-6199}}$,$^{2,4}$\thanks{E-mail: elina.lindfors@utu.fi}
Kari Nilsson,$^{2}$
Pekka Hein\"am\"aki,$^{5}$
Jari Kotilainen$^{2,4}$
\\
$^{1}$Department of Physics, Norwegian University of Science and Technology, NO-7491 Trondheim, Norway\\
$^{2}$Finnish Centre for Astronomy with ESO (FINCA), University of Turku, V\"ais\"al\"antie 20, 21500 Piikki\"o, Finland \\
$^{3}$Aalto University Mets\"ahovi Radio Observatory, PO Box 13000, FI-00076 Aalto, Finland \\
$^{4}$Department of Physics and Astronomy, Vesilinnantie 5, University of Turku, 20014 Turku, Finland \\
$^{5}$Tuorla Observatory, Department of Physics and Astronomy, Vesilinnantie 5, University of Turku, 20014 Turku, Finland
}

\date{Accepted XXX. Received YYY; in original form ZZZ}

\pubyear{2015}

\begin{document}
\label{firstpage}
\pagerange{\pageref{firstpage}--\pageref{lastpage}}
\maketitle

\begin{abstract}
Blazars constitute the most numerous source class in the known extragalactic population of very high energy (VHE) gamma-ray sources. However, determining their redshifts is often challenging due to weak or non-existent emission lines in their spectra. This study focuses on two BL Lacs, KUV 00311-1938 and S2 0109+22, where previous attempts at redshift determination have faced difficulties. By combining spectroscopic observations with photometric redshift estimates, we tentatively assign a redshift of $z=0.634$ to KUV~00311-1938 and a likely redshift of $z=0.49$ to S2 0109+22. Establishing redshift estimates for high-redshift blazars is crucial for understanding extragalactic VHE gamma-ray sources and their interactions with the surrounding universe.  
\end{abstract}




\section{Introduction}

The number of known extragalactic very high energy (VHE) gamma-ray sources, with photon energies above 100 GeV, has increased from 10 to approximately 100 in the last 15 years, a trend anticipated to accelerate with the advent of the Cherenkov Telescope Array (CTA)\footnote{\url{https://www.cta-observatory.org}}. These sources are predominantly blazars, a subset of Active Galactic Nuclei (AGN) characterized by a relativistic jet aligned closely with our line of sight. Consequently, the jet's emission experiences strong Doppler boosting. 

Nearly 80\% of the VHE gamma-ray-detected blazars belong to a group called BL Lacertae objects (BL Lacs), characterized by intrinsically weak optical emission lines likely resulting from a radiatively inefficient accretion flow toward the central supermassive black hole \citep{rees82}. Determining their redshifts through spectroscopy is particularly challenging due to the absence of emission lines that are overshadowed by the non-thermal continuum obscuring host galaxy features. However, fairly accurate redshift estimates can still be obtained based on absorption lines/edges of the H I Lyman $\alpha$ forest \citep[e.g.][]{Danforth10,dorigojones22}, redshift lower limits from intervening absorption systems \citep[e.g.][]{furniss13,2014A&A...565A..12P}, and host galaxy CO lines in the millimeter regime \citep{fumagalli12}, which do not depend on the power of the non-thermal continuum. Nevertheless, many BL Lacs ($\sim$18\%) still lack confirmed redshifts, despite numerous major observing programs targeting this abundant extragalactic VHE gamma-ray source population \citep[e.g.][]{Paiano2020,Goldoni2021, Kasai2023}. Moreover, the leading candidate source of astrophysical neutrinos, TXS 0506+056 \citep{IceCube2018}, is a BL Lac object with weak spectral lines \citep{Paiano2017}, further emphasizing the need for redshift determination efforts \citep{Paiano2021}.

BL Lacs, as bright extragalactic sources, provide valuable insights into the region of the universe between us and the blazar. VHE gamma-rays are absorbed through interactions with near- and mid-infrared photons of the extragalactic background light (EBL) that permeates the universe. This absorption is energy-dependent and becomes more pronounced with increasing redshift. Prior to modeling the VHE emission of blazars, this distortion must be corrected, requiring knowledge of the redshift of the source. Conversely, if the intrinsic blazar spectrum is known, VHE observations can be used to constrain the EBL in the infrared-UV bands. Thus, blazars at different redshifts can serve as indicators of the EBL density at varying wavelengths. The most recent limit, provided by \citet{acciari19} based on results from 32 gamma-ray spectra for 12 blazars within the redshift range of $z=0.03-0.944$, is consistent with predictions from current EBL models. While the CTA has the potential to enhance these limits \citep{CTA_gpropa}, achieving this goal requires an increase in the number of blazars with known redshifts, particularly those with $z>0.4$, for which we currently have only ten sources\footnote{see e.g. \url{http://tevcat.uchicago.edu/}}. In addition to studying the EBL, X-ray bright BL Lacs can also be utilized to investigate the warm-hot intergalactic medium \citep[WHIM, log $T(K)=5-7$; e.g.][]{fang2007}.

In this study, we aim to determine the redshifts of two BL Lacs: KUV~00311-1938 and S2~0109+22. Direct spectroscopic redshift determinations for these sources have proven challenging, and they are likely to be located within a redshift range where blazars are scarce for indirect EBL studies. To determine the likely redshift of the blazars, we use multi-object spectroscopy to search for possible cosmic neighbours by assessing the redshifts of the galaxies in the field of view and comparing them with the previous redshift limits from direct spectroscopy and deep imaging of the blazars. A similar method has been applied to other blazars previously: PKS 0447-439 ($z=0.343$; \citealt{Muriel15}), PKS 1424+240 ($z=0.601$; \citealt{Rovero16}), PKS 2155-304 ($z=0.116$; \citealt{farina16}), 3C~66A ($z=0.34$; \citealt{2018MNRAS.474.3162T}), PG 1553+113 ($z=0.433$; \citealt{Johnson19}), RGB 2243+203 ($z=0.528$; \citealt{2019MNRAS.482.5422R}), and S5 0716+714 ($z=0.23$; \citealt{pichel23}).

KUV~00311-1938 was suggested to be one of the farthest BL Lacs detected at VHE, based on a tentative redshift of $z=0.61$ by \citet{piranomonte07}. However, neither \cite{2014A&A...565A..12P} nor \cite{pichel21} could confirm this redshift. \cite{2014A&A...565A..12P} clearly identified a Mg~II doublet corresponding to a redshift of $z=0.506$, providing a lower limit for the distance of KUV~00311-1938. \cite{pichel21} also presented observations of 41 galaxies around this BL Lac, but due to the absence of a numerous group of galaxies, they could not conclusively determine the redshift.

For S2~0109+22, \cite{Paiano2016} derived a redshift of $z>0.35$ through optical spectroscopy. However, they also reported the presence of a group of faint galaxies at $z\sim0.26$, adopting photometric redshifts from the Sloan Digital Sky Survey (SDSS). The detection of the faint host galaxy in the near-IR band, as reported in \cite{2018MNRAS.480..879M}, resulted in a redshift estimation of $z=0.36\pm0.07$.

In this paper, we identified cosmic neighbouring galaxies around these two blazars, using both spectroscopic observations obtained from the European Southern Observatory's Very Large Telescope (VLT), in conjunction with previous observations from the literature (see Section 2), and photometric redshifts of the nearby galaxies derived from SDSS Data Release 17 (SDSS/DR17; see Section 3). Associating these cosmic neighbours with the blazars, combined with redshift constraints from previous observations, enabled us to determine a redshift of $z=0.49$ for S2~0109+22 and a tentative redshift of $z=0.64$ for KUV~00311-1938. We use a flat cosmology with $H_0 = 69.6$ km s$^{-1}$ Mpc$^{-1}$ for our calculations.

\section{Spectroscopic observations and analysis}

\begin{figure*}
 \centering
 \includegraphics[width=0.95\linewidth]{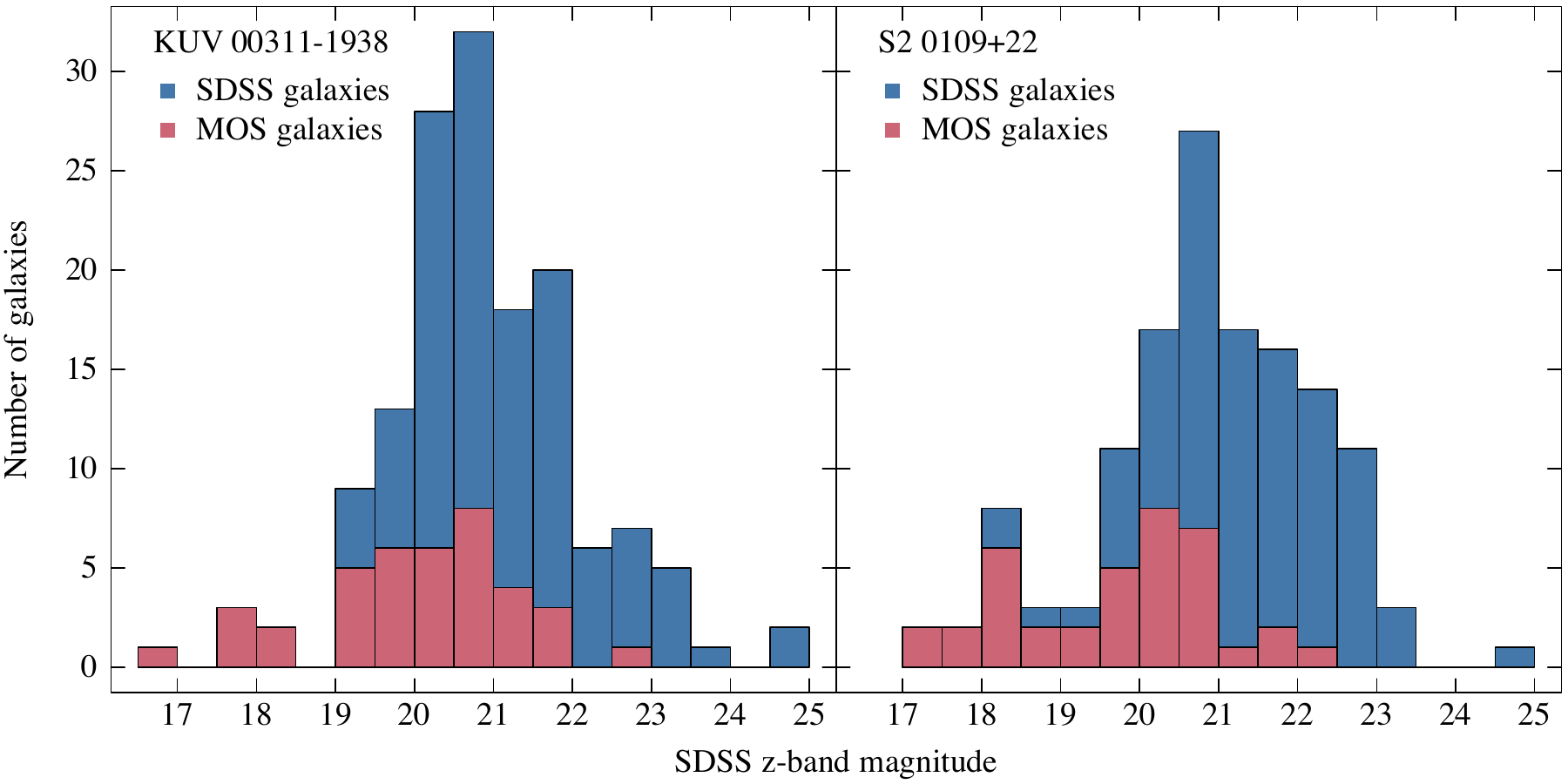}
 \caption{SDSS z-band magnitude histograms of the sources located in the fields of view of the FORS2 observations shown in Figs. \ref{fig:kuv_mosaic} and \ref{fig:S2_mosaic}. Sources probed by the MOS observations studied in this paper are highlighted in red, while all identified SDSS galaxies are highlighted in blue.
 }
 \label{fig:zmag_histogram}
\end{figure*} 

We conducted observations of the galaxies located close to KUV~00311-1938 and S2~0109+22 using the VLT's FOcal Reducer/low dispersion Spectrograph 2 (FORS2) instrument in the Multi-Object Spectrometer (MOS) mode. Additionally, we analyzed archival data from the International Gemini Observatory for both fields, acquired with the Gemini Multi Object Spectrograph (GMOS) instrument. In the following sections, we describe the data reduction processes and present the full galaxy sample, including spectroscopic and photometric redshift estimates for additional galaxies in the FORS2 fields of view, as obtained from the literature and SDSS/DR17.

\subsection{FORS2}

We conducted observations of 24 targets in the vicinity of KUV~00311-1938 and 30 targets in the vicinity of S2~0109+22 using FORS2/MOS on 22--23 October 2017 (P100). The selection of targets was based on their brightness and extended structure, considering the constraints imposed by the slit arrangement and available observing time. As a result, primarily the brightest targets in the field were chosen. The ESO grism 600RI was employed, providing a wavelength range of 512--845 nm for the resulting spectra. The FORS2 chips were exposed for 600 seconds each, and this process was repeated three times. The airmass ranged from 1.1 to 1.2 for the field of KUV~00311-1938 and 1.5 to 1.6 for S2~0109+22. On October 22, 2017, the seeing ranged from 0.8 arcsec to 1.4 arcsec during the observing blocks, whereas on October 23, 2017, it varied between 0.7 arcsec and 0.9 arcsec. We processed the FORS2/MOS data using FORS pipeline release 5.5.6 in EsoRex \citep{freudling13}. In both fields, we utilized the same spectroscopic standard; HILT 600. We successfully processed the spectra from 21 and 15 galaxies in the field of KUV~00311-1938 and S2~0109+22, respectively. The remainder were either too faint or identified as stars.

\subsection{GMOS}

We retrieved the International Gemini Observatory data (observing programs GS-2016B-Q-55 and GN-2017B-Q-50; PI Pichel) from the Gemini Observatory Archive\footnote{\url{https://archive.gemini.edu}}. These data were acquired using GMOS on October 5, 2016, in the field-of-view of approximately $5 \times 5$ arcmin$^2$ centered on the blazar KUV~00311-1938. Additional data were obtained on September 27, 2017, for the field of the blazar S2~0109+22. The KUV~00311-1938 field data were obtained using the Gemini South telescope, while S20109+22 field data were acquired using the Gemini North telescope. In both cases, the exposure time for the field was 5$\times$900s, divided for different dispersion angles (ranging from 590 nm to 630 nm), and the grating used was B600+G5323, providing a wavelength range of approximately 400--700 nm for the resulting spectra. For the KUV~00311-1938 field, the airmass ranged from 1.0 to 1.1, and seeing from 0.7 arcsec to 1.3 arcsec, with the spectroscopic standard employed being LTT 7379. In contrast, for the S2~0109+22 field, the airmass varied between 1.0 and 1.1, with seeing approximately $\sim$0.9 arcsec, and the spectroscopic standard used was G191B2B. We processed the Gemini data using \textsc{PypeIt} \citep{pypeit:zenodo,pypeit:joss_arXiv,pypeit:joss_pub}. Out of 43 and 39 targets for the fields of KUV~00311-1938 and S2~0109+22, respectively, we successfully obtained new spectra from 11 and 23 galaxies, with the remaining targets being either too faint, acquisition targets, stars, identical to those in the FORS2 data, or located outside the FORS2 fields of view.

\subsection{The full galaxy sample}

For all sources, we averaged spectra from different exposures. In cases with a low signal, we also binned them by a factor of up to four. We estimated source redshifts by scaling a galaxy template spectrum, selected from five galaxy templates\footnote{Obtained from \url{https://classic.sdss.org/dr5/algorithms/spectemplates}} ranging from early- to late-type, to the normalization of a given observed spectrum and shifting the wavelength to align with it.

Additionally, we complemented our sample with redshift estimates of the galaxies in the field of KUV~00311-1938 from \citet{pichel21}, who used data from Gran Telescopio Canarias (GTC) and the MOS instrument Optical System for Imaging and low-Intermediate-Resolution Integrated Spectroscopy (OSIRIS). In total, we obtained redshift information for 39 and 38 galaxies in the fields of KUV~00311-1938 and S2~0109+22, respectively.

We also obtained photometric redshifts for all galaxies in both fields from SDSS/DR17 \citep{abdurrouf22} and compared them to spectroscopic redshifts obtained in our work and previous studies. We noticed that a few redshifts derived in earlier studies deviated significantly ($>3\sigma$) from the photometric redshift estimates. Comparing our derived redshifts from Gemini data to those reported by \citet{pichel20,pichel21}, we observed agreement for all galaxies except for seven (slits 7, 19, 20, 24, 27, and 29 for the field of S2~0109+22, and slit 29 for the field of KUV~00311-1938). In this paper, we use the results of our new analysis throughout.

Fig. \ref{fig:zmag_histogram} displays the SDSS z-band magnitude histograms of both fields, with the VLT, Gemini, and GTC targets highlighted in red, alongside all identified galaxies in the SDSS/DR17 shown in blue. It is evident that our targets do not constitute a complete flux or volume limited sample.

\section{Results and discussion}

\begin{figure*}
 \centering
 \includegraphics[clip,trim={280 400 380 260},width=0.95\linewidth]{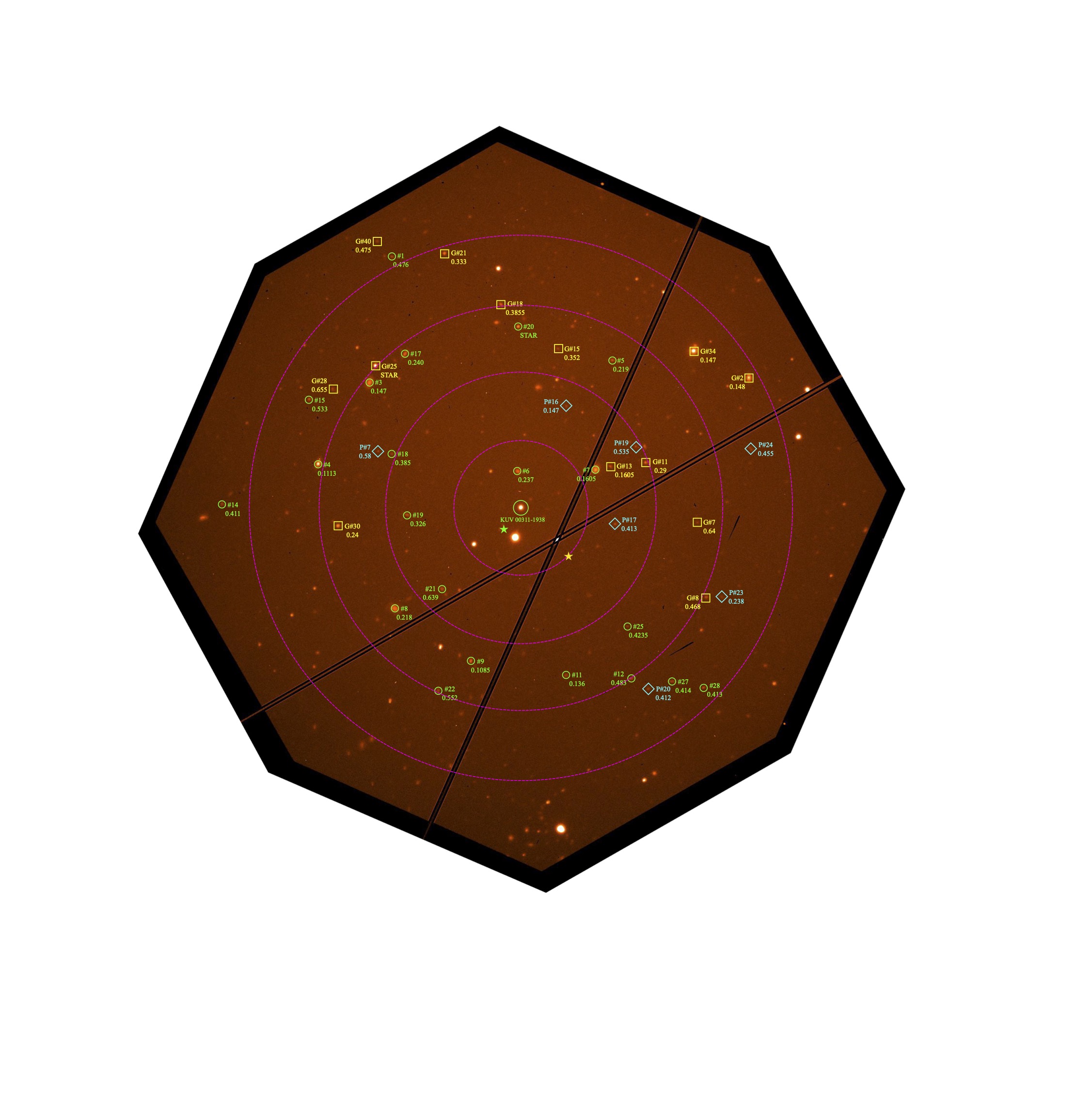}
 \caption{Mosaic R-band image of the field of KUV 00311-1938 taken with FORS2 from our observations. The green circles indicate the sources observed with FORS2 MOS, the yellow squares indicate the sources observed with Gemini, and the blue diamonds mark the GTC sources from \citet{pichel21}. Corresponding source redshifts are shown along with the source numbering. Sources with a stellar spectrum are marked as `STAR'. Magenta dashed lines show angular radii of 0.5, 1.0, 1.5, and 2.0 Mpc at a redshift of $z=0.64$ (the most likely distance of the blazar based on our analysis). The yellow and green stars show the center locations of the spectroscopic and photometric groups at $z=0.64$ and $z=0.7$, respectively, derived from our analysis. The black lines crossing the image come from chip gaps and the image is cropped to show only the area covered by the two exposures.
 }
 \label{fig:kuv_mosaic}
\end{figure*} 

\begin{figure*}
 \centering
 \includegraphics[width=0.95\linewidth]{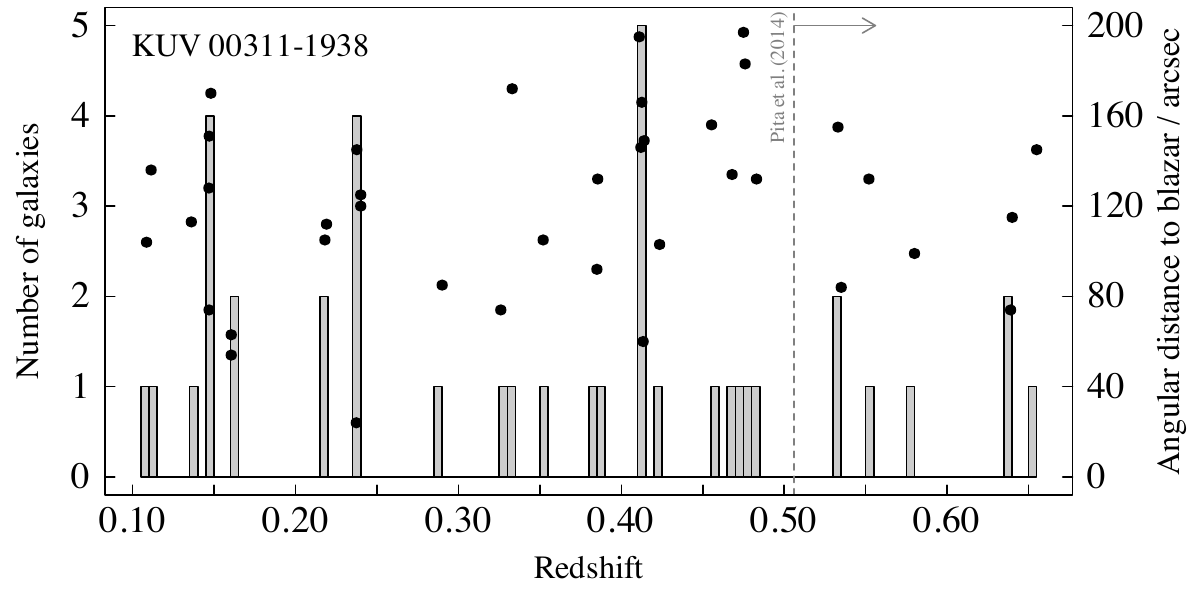}
 \caption{Redshift histogram of all galaxies studied in this paper in the field of KUV 00311-1938 (see Fig. \ref{fig:kuv_mosaic}). The individual galaxy redshifts are also plotted as points, indicating their angular distance from the blazar position. The field contains two galaxy groups with four members each and one with five. A distance lower limit from the literature is shown as a dashed line.   
 }
 \label{fig:kuv_histogram}
\end{figure*} 

\begin{figure*}
 \centering
 \includegraphics[clip,trim={330 450 390 280},width=0.95\linewidth]{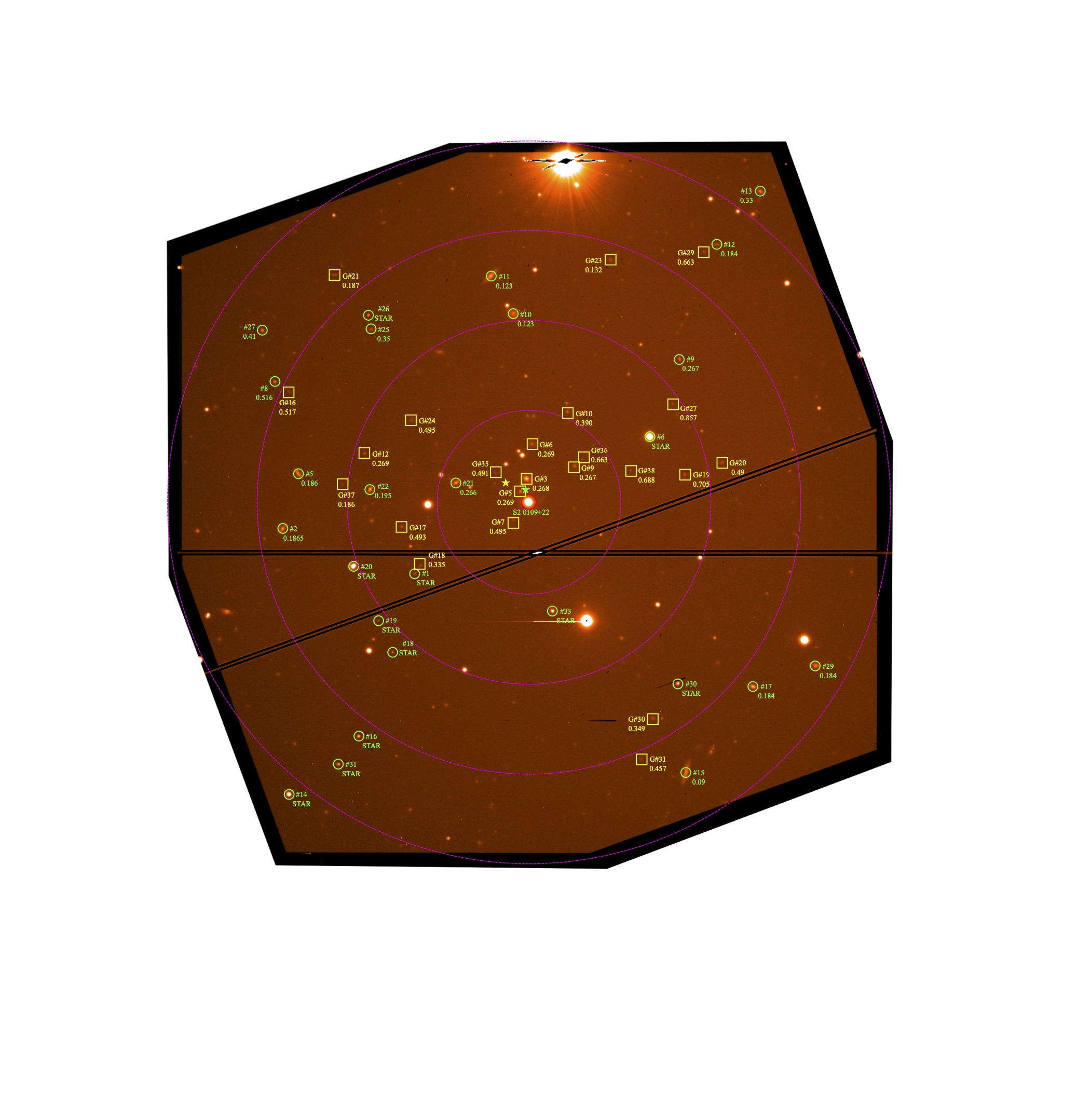}
 \caption{Mosaic R-band image for S2~0109+22 taken with FORS2 from our observations. The green circles indicate the sources observed with FORS2 MOS, and the yellow squares indicate the sources observed with Gemini. Corresponding source redshifts are shown along with the source numbering. Sources with a stellar spectrum are marked as `STAR'. Magenta dashed lines show angular radii of 0.5, 1.0, 1.5, and 2.0 Mpc at a redshift of $z=0.49$ (the most likely distance of the blazar based on our analysis). The yellow and green stars show the center locations of the spectroscopic and photometric groups at $z=0.49$, respectively, derived from our analysis. The black lines crossing the image come from chip gaps and the image is cropped to show only the area covered by the two exposures.  
 }
 \label{fig:S2_mosaic}
\end{figure*} 

\begin{figure*}
 \centering
 \includegraphics[width=0.95\linewidth]{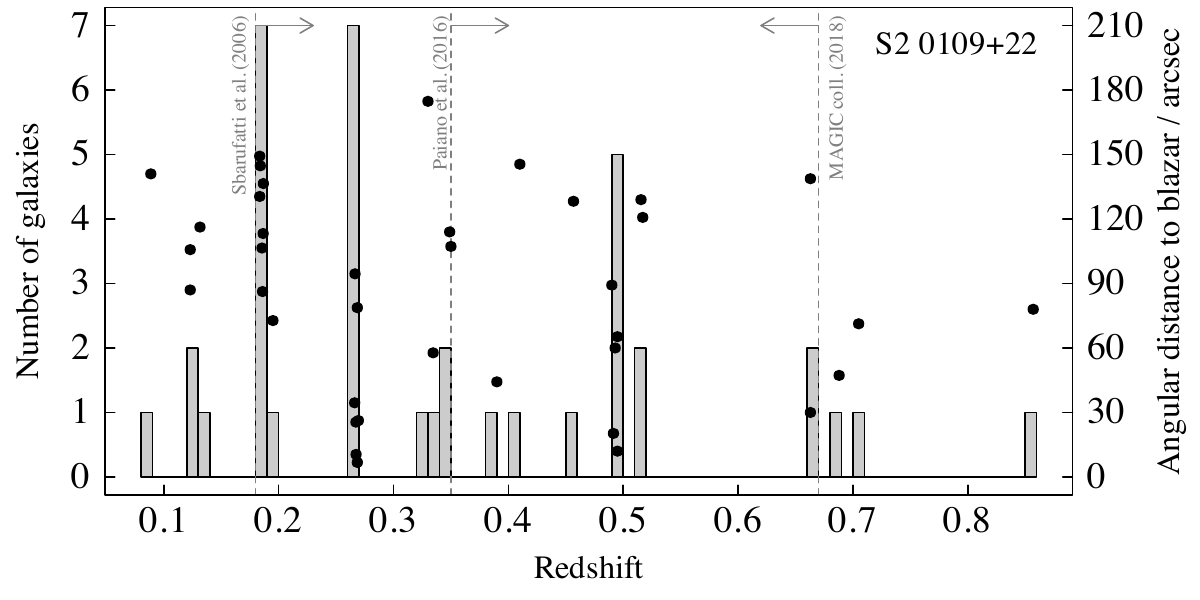}
 \caption{Redshift histogram of all galaxies studied in this paper in the field of S2~0109+22 (see Fig. \ref{fig:S2_mosaic}). The individual galaxy redshifts are also plotted as points, indicating their angular distance from the blazar position. The field contains three galaxy groups with either five or seven members. Three distance lower/upper limits from the literature are shown as dashed lines.  
 }
 \label{fig:S2_histogram}
\end{figure*} 

The estimated spectroscopic redshifts of all galaxies used in this work are presented in Figs.~\ref{fig:kuv_mosaic}--\ref{fig:S2_histogram}, displayed as mosaic images and histograms to identify potential groups of galaxies sharing similar redshifts. It is important to note that these identified groups of cosmic neighbors may not necessarily be genuine galaxy groups; they could also include galaxies from neighboring groups or be part of the same cluster of galaxies. This ambiguity arises because, in this redshift range, we only observe the brightest galaxies in the groups. For simplicity, however, we refer to them as `group candidates' in the following sections. Detailed redshift data for the galaxies can be found in Tables \ref{tab:kuv_table} and \ref{tab:s2_table}. 

By combining our FORS2 data with reanalyzed Gemini data and incorporating GTC galaxies from \cite{pichel21}, we have identified multiple galaxy group candidates in the field of KUV~00311-1938. Specifically, there are two group candidates with four members each, located at $z=0.147-0.148$, and $z=0.237-0.240$, respectively. Additionally, there is one group candidate with five members at $z=0.411-0.414$. Furthermore, two galaxy pairs are found at $z=0.533-0.535$, and $z=0.639-0.640$. Similarly, in the field of S2~0109+22, we have identified two group candidates, each with seven members, located at $z=0.1835-0.1867$ and $z=0.2662-0.2694$, respectively. Additionally, there is one group candidate with five members at $z=0.490-0.495$.

\subsection{Comparison with previous redshift limits}

Previous redshift limits for both sources were determined through direct optical spectroscopy and Very High Energy (VHE) $\gamma$-ray observations. Lower limits from optical spectroscopy are derived either by detecting lines from intervening absorption systems \citep{2014A&A...565A..12P} or, in cases where no lines are detected, by assuming that the host galaxy of the blazar is a giant elliptical galaxy and estimating how faint the galaxy must be for its lines not to be detected in the obtained spectrum \citep{sbarufatti,Paiano2016}. On the other hand, upper limits for redshifts from the $\gamma$-ray spectrum, based on EBL absorption arguments, have also been established for both sources. \cite{hess_kuv} set an upper limit of $z<0.98$ for KUV~00311-1938, while \citet{2018MNRAS.480..879M} reported an upper limit of $z\leq0.67$ for S2~0109+22. These upper limits exceed the redshifts of the group candidates we detected in the fields of these two blazars, thus confirming all identified group candidates as potential associations with the blazar. However, in the following, we will compare the spectroscopic lower limits with the redshifts of the group candidates identified in our analyses.

In the case of KUV~00311-1938, \cite{2014A&A...565A..12P} obtained an X-Shooter spectrum, and the UVB arm spectrum revealed a Mg~II doublet at 4215\AA, corresponding to a redshift of $z=0.506$. This serves as a strict lower limit to the blazar's redshift. This finding makes the association with the most numerous group candidates at $z=0.147-0.148$, $z=0.237-0.240$, and $z=0.411-0.414$ unfeasible. No companion galaxies were detected at the distance of $z\sim0.5$. If the absorption system is intrinsic to the host galaxy, this suggests that KUV~00311-1938 is isolated at this distance. However, since our sample does not encompass all galaxies in the field, we might have missed the ones at this particular distance. In addition, there could be an intervening cold system absorbing the blazar light along the line-of-sight at $z=0.506$, but the blazar itself could be positioned farther away. We identified two galaxy pairs situated at $z=0.533-0.535$ and $z=0.639-0.640$, which could potentially be cosmic neighbours of the blazar. We will further evaluate this possibility in the following sections.

For S2~0109+22, \cite{Paiano2016} presented a GTC spectrum of the blazar, which did not reveal discernible line features. They simulated source spectra by assuming an absolute magnitude for the host galaxy of M$_{\rm R,Vega}$ = -22.9 mag. If the blazar were associated with the group candidate at $z=0.265$, Ca~II and G-band lines would have been visible in the GTC spectrum. Consequently, they suggested a redshift of $z>0.35$ based on the assumed brightness of the host galaxy.
 
We repeated this simulation analysis at three different redshifts: $z=0.18$, $z=0.26$, and $z=0.49$, as determined by our spectroscopic galaxy group candidate analysis. In each case, the apparent R-band magnitude of S2~0109+22 was assumed to be m$_{\rm R,Vega}$ = 14.8 mag, as reported in \citet{Paiano2016}. Consequently, the core-to-host flux ratio increased from 5.9 to 80 as the redshift increased from $z=0.18$ to $z=0.49$. We assumed a magnitude of M$_{\rm R,Vega}$ = -22.9 mag and an effective radius of 9 kpc for the host galaxy, applied the K- and evolution corrections, and accounted for the slit losses, which varied with redshift for both the core and the host galaxy. While \citet{Paiano2016} did not specify the full width at half maximum (FWHM) of their observation, assuming FWHM = 0.7 arcsec provided a very accurate reproduction of the correct flux level. Finally, we introduced Gaussian noise with the same characteristics as in the observed spectrum into the simulation.
 
\begin{figure*}
 \centering
 \includegraphics[width=0.95\linewidth]{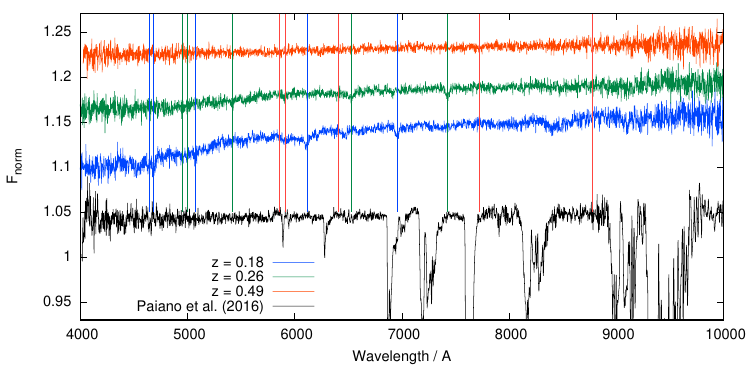}
 \caption{Simulated spectra of S2~0109+22 at the three redshifts discussed in the text and the spectrum from \citet{Paiano2016}. We assumed a magnitude of M$_{\rm R,Vega}$ = $-$22.9 mag for the host galaxy. All spectra have been normalized by a power-law continuum with a slope -0.56. The spectra have been shifted vertically for clarity. The vertical lines indicate the positions of the (from left to right) Ca K, Ca H, G-band, Mg b, and Na D lines with the same color coding as the spectra.}
 \label{fig:paiano_comparison}
\end{figure*} 

Figure \ref{fig:paiano_comparison} illustrates the results of the simulations. A visual inspection of the spectra clearly indicates that the host galaxy absorption lines would certainly be detectable at $z=0.18$ and very likely at $z=0.26$. In contrast, at $z=0.49$, they would be impossible to detect. This supports the conclusion by
\cite{Paiano2016} that the redshift is likely to be larger than $z>0.35$, suggesting that the most numerous candidate groups in the field are likely to be foreground groups.

We previously conducted deep imaging of S2~0109+22 and detected the host galaxy with an apparent I-band magnitude of m$_{\rm I,Vega}$ = 18.05 mag \citep{2018MNRAS.480..879M}. In that study, assuming the host galaxy to be a giant elliptical galaxy with M$_{\rm R,Vega}$ = -22.8$\pm$0.5 mag, we utilized the method of \citet{sbarufatti05} to derive an imaging redshift of $z=0.36\pm0.07$. The association with a more distant candidate group suggests that the host galaxy is somewhat more luminous, M$_{\rm R,Vega}$ = -23.8$\pm$0.5 mag, than typical blazar host galaxies. This aligns with our findings in \citet{Nilsson03}: while imaging redshifts agree very well at redshifts $z<0.3$, at higher redshifts, we tend to detect only the most luminous host galaxies, making the imaging redshifts less accurate.

\subsection{The probability of observing galaxy groups by chance}

To assess the probability that the observed galaxy group candidates occur in the fields by chance, we followed the methodology outlined in \citet{Rovero16}. We utilized the galaxy group catalogue, zCOSMOS\footnote{\url{https://github.com/wkcosmology/zCOSMOS-bright\_group\_catalog}}, containing groups within the redshift range of $0.1<z<1.0$, and an apparent magnitude range of m$_{\rm I,AB}<22.5$ mag, to randomly select positions in the sky. This process was based on the assumption that the sky exhibits similar characteristics throughout, as the zCOSMOS field is relatively small and not positioned in the same location as the source. In addition, since our sample is not complete to m$_{\rm I,AB}$ $\approx$ m$_{\rm z,AB}$ $<$ 22.5 mag, this estimation is conservative, as we are very likely missing members from our candidate groups. This enabled us to determine the likelihood of a coincidental presence of groups with a specified minimum number in the observed field. The resulting coincidence probabilities are detailed in Table \ref{tab:zcosmos} for both fields, various redshift ranges, and sizes of galaxy groups.

\begin{table}
    \caption{Chance coincidence probabilities for several parameter sets using the galaxy group catalogue from zCOSMOS.} 
    \centering
    \begin{tabular}{c|ccc}   
    \hline
    Redshift & N & prob & prob$\times$0.3 \\
    \hline
    \textbf{KUV-like params:} \\
    $0.50<z<0.65$ & $\geq2$ & 0.56 & 0.17 \\
                  & $\geq3$ & 0.22 & 0.066 \\
    \hline
    \textbf{S2-like params:} \\
    $0.49<z<0.85$ & $\geq5$ & 0.18 & 0.054 \\
                  & $\geq6$ & 0.12 & 0.036 \\
    \end{tabular}
    \label{tab:zcosmos}
\end{table}

Furthermore, assessing the distance between galaxies within galaxy groups and the blazar can help estimate the likelihood of association. Galaxy groups typically have sizes smaller than 2\,Mpc \citep[see e.g.][]{Tago10}, making this a common search radius for cosmologically neighboring galaxies.
\citet{Massaro19a,Massaro19b,Massaro20} investigated groups around radio galaxies and BL Lac objects, revealing that if there are two companions at the same redshift ($z<0.005$) within a distance of 1.0 Mpc from the blazar and/or four companions within 2.0 Mpc, the probability of the blazar being associated with that group exceeds 95\%. In addition to distance, the location within the group is important. \citet{Massaro19a,Massaro19b,Massaro20} noted that blazars tend to be positioned at the group centers. Utilizing Galaxy and Mass assembly data at higher redshifts ($0.1<z<0.35$), \cite{Wethers22} observed a similar tendency for quasars in general. Although these studies do not cover the redshift range of our candidate groups and may not be directly applicable to our objects, we nonetheless explored the virial radii of the candidate groups and the position of the blazar within each candidate group.

For KUV~00311-1938, the estimated virial radii of the candidate group, which includes the blazar and the pair of galaxies at $z=0.533-0.535$ or $z=0.639-0.640$, are 1.25\,Mpc and 1.21\,Mpc, respectively. These values are notably larger than typical galaxy group sizes \citep[see e.g.][]{Nurmi2013}. Therefore, instead of constituting galaxy groups, they are likely neighbouring groups, and we observe only the brightest galaxies from these structures. Additionally, we computed the geometrical center for the pairs, and in both cases, they are within a few hundred kiloparsecs from the blazar. The central point for the group at $z=0.639-0.640$ is marked by a green star in Fig. \ref{fig:kuv_mosaic}.

In the case of S2~0109+22, there are five galaxies in the candidate group at $z=0.49$ that may be associated with the blazar. Assuming the blazar belongs to this group, we derive a virial radius of 0.35\,Mpc, falling within the range derived for luminous groups in \cite{Nurmi2013}. Additionally, the geometrical center of this candidate group is very close to the blazar (see Fig. \ref{fig:S2_mosaic}).

\subsection{Search for possible groups using photometric redshifts}

\begin{figure*}
 \centering
 \includegraphics[width=0.95\linewidth]{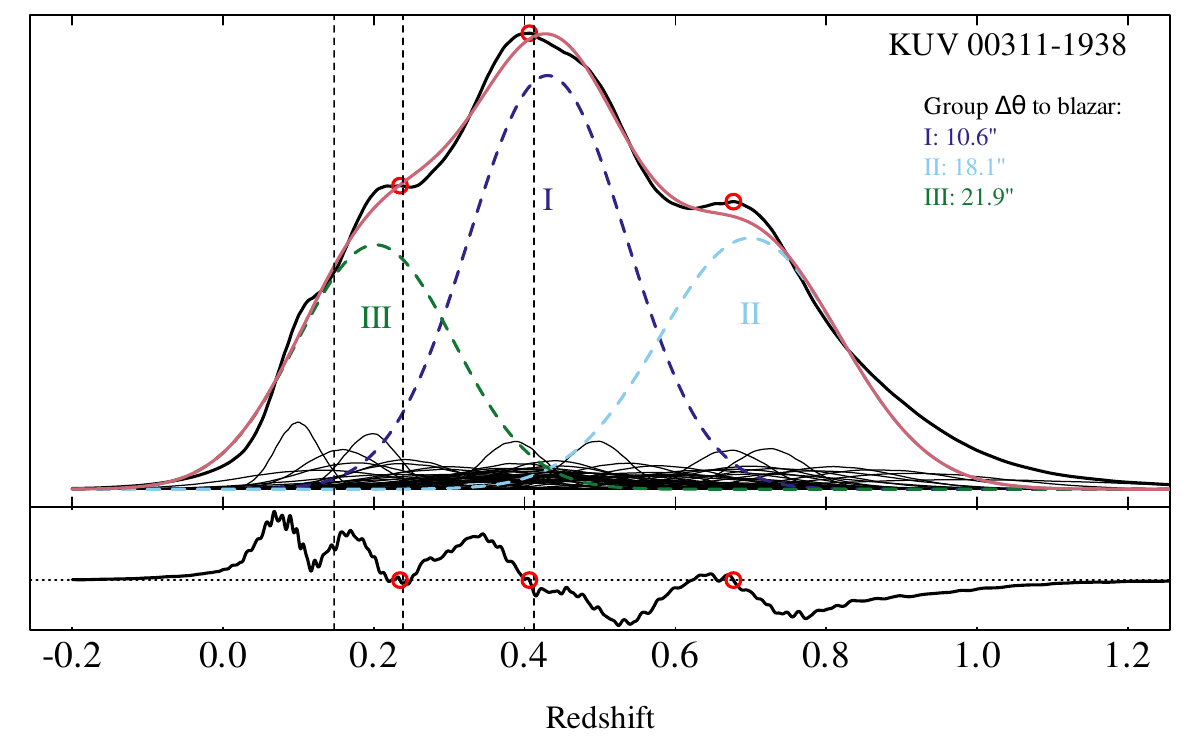}
 \caption{A KDE analysis for the SDSS/DR17 photometric redshift distributions of galaxies within 2' from KUV~00311-1938. \textit{Upper panel:} The solid thin black lines show the KDEs for single galaxies in the field; the thick solid black line is their sum. The solid red line shows the estimated galaxy distribution in the field assuming that it consists mostly of a few individual galaxy groups (in this case three; colored dashed lines and marked with roman numerals). The average, weighted group center separation to the location of the blazar is marked in the figure legend. \textit{Lower panel:} Local derivative of the summed KDE. Red circles show the crossing points of the derivative from positive to negative values marking the locations of local maxima in the summed KDE (shown also in the upper panel). Black vertical dashed lines show the locations of the galaxy groups in the field from our spectral analysis. They match well with the zero crossings and peaks of the individual Gaussian profiles (I and III). The KDE analysis suggests an additional galaxy group in the field at z$\sim$0.68 (II) not picked up by the spectral analysis that could be associated with the blazar.   
 }
 \label{fig:photoZ_kuv}
\end{figure*}

\begin{figure*}
 \centering
 \includegraphics[width=0.95\linewidth]{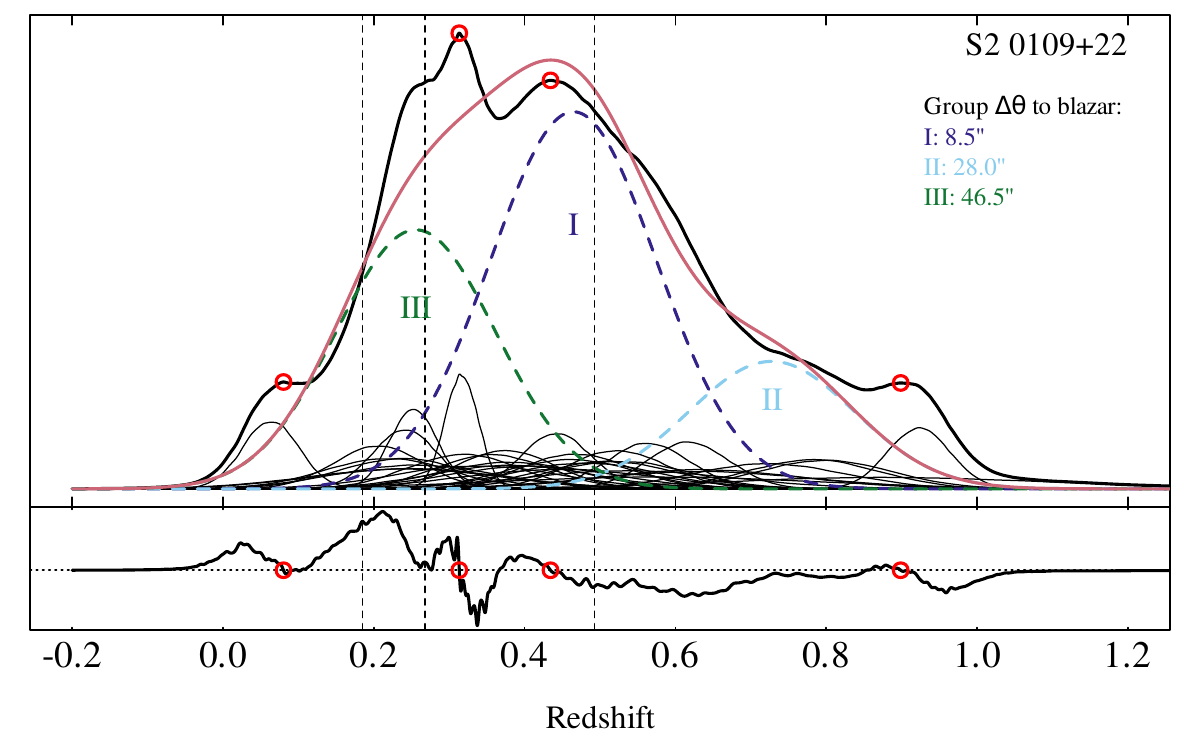}
 \caption{A KDE analysis for the SDSS/DR17 photometric redshift distributions of galaxies within 2' from S2~0109$+$22 similar to Figure \ref{fig:photoZ_kuv}. The peaks of the individual Gaussian profiles (I and III) match well with the spectrometric redshifts, while the local maxima are dominated by KDEs from single sources. The average weighed group centers suggest group I as the most probable association to the blazar.  
 }
 \label{fig:photoZ_s2}
\end{figure*}

We obtained the photometric redshifts for galaxies within our fields of view using SDSS/DR17, resulting in 52 and 36 redshift estimates for the fields of KUV~00311-1938 and S2~0109+22, respectively. To assess potential galaxy group candidates based on the photometric redshift data, we created kernel density estimates (KDEs) of the redshift distributions, assuming a Gaussian distribution with a centroid and width corresponding to the values tabulated in SDSS/DR17. The corresponding KDEs, including a combined one, are depicted in Fig. \ref{fig:photoZ_kuv} and Fig. \ref{fig:photoZ_s2} for both fields.

Assuming that the combined KDE is primarily composed of a few galaxy groups (neglecting lower-level `noise' from individual galaxies situated at random distances), we can approximate the combined KDEs using a few Gaussian components (colored dashed lines in Figs. \ref{fig:photoZ_kuv} and \ref{fig:photoZ_s2}) representing the individual groups. These Gaussian components have centroids at the group redshift and widths similar to the mean error of the photometric redshift values in SDSS/DR17 ($\Delta z \sim 0.12$). We note again that, for simplicity, we refer to these groupings as `group candidates', but they are more likely to trace some larger structure such as neighbouring groups or a cluster.

For KUV~00311-1938, we fitted the combined KDE with three Gaussian components. These three Gaussians, representing three galaxy group candidates, adequately account for the overall galaxy distribution. The Gaussian centroids of the two lower redshift groups closely align with the redshifts of the group candidates obtained from the multi-object spectroscopy, considering that the group candidates at $z=0.18$ and $z=0.27$ are likely blended with each other. Additionally, a third group candidate at $z\sim0.68$ is necessary to explain the combined KDE at higher redshifts. This peak is sufficiently prominent in the field, suggesting it may represent a potential galaxy group located approximately at that distance, which could be associated with the blazar. In the spectral analysis, we identified a galaxy pair at a redshift of $z=0.639-0.640$ that could represent galaxies in this group candidate, favoring this redshift over the other galaxy pair at z$=0.533-0.535$. Nevertheless, this remains a tentative association, and further data would be required to confirm it.

Similar to KUV~00311-1938, the galaxy distribution in the field of S2~0109+22 can be represented by three Gaussian components. Their centroid values again closely match those obtained through our spectroscopic analysis. However, the third group candidate identified in our spectral analysis at $z=0.19$ is not clearly discernible in the combined KDE, likely because the redshift is quite close to the group at $z=0.27$. Thus, the SDSS photometric redshifts of galaxies in the field of S2~0109+22 support the existence of the three group candidates identified in our spectral analysis.

Finally, we calculated the `composite' group central location by weighting the coordinates (RA/Dec) of each galaxy according to the Gaussian distribution of a given group and based on the individual galaxy redshifts (i.e., the weight would be unity if the redshift of a galaxy matches the peak redshift of the Gaussian component and gradually diminishes towards the tails of the distribution). As mentioned above, it is reasonable to assume that blazars should be situated near the center of their associated group candidates \citep{Massaro19a,Massaro19b}. Therefore, the closer the group candidate center is to the blazar, the more likely the association. We computed the geometrical center of the candidate groups in both fields, and their separation from the blazar is marked in Figs. \ref{fig:photoZ_kuv} and \ref{fig:photoZ_s2}. In the field of KUV~00311-1938, the weighted group centers are roughly equally proximate to the blazar. However, in the field of S2~0109+22, the weighted group center close to the redshift of $z=0.49$ (group I in Fig. \ref{fig:photoZ_s2}) is notably much closer than the other two, further strengthening its association with the blazar.   

\subsection{The probability that a blazar is not associated with any group}

It is possible that the blazar is isolated and not associated with any group that may be present in the field. However, there are limited studies to accurately quantify this. \citet{muriel16} analyzed a sample of approximately 120 blazars and determined that 32$\pm$4\% of them exist as single sources, while 43$\pm$5\% are found in groups containing three or more members. Nevertheless, \citet{Massaro19b} critiqued the sample used in this study, concluding that only 14 sources of the sample were genuinely blazars. Among these 14, four were discovered to be isolated (i.e., 29\%), and seven were found in groups of three or more members (i.e., 50\%), aligning with \citet{muriel16}. It is worth noting that the blazars examined in these studies have redshifts less than $z<0.2$. Based on the above, it appears that approximately 30\% of blazars tend to be isolated. This probability is independent of the likelihood of finding groups in a random position in the sky. Consequently, the combined probabilities of having a non-associated galaxy group in the same field with the blazar are tabulated in Table \ref{tab:zcosmos}. Essentially, the probability that the blazar S2~0109+22 is isolated and the field coincidentally hosts a galaxy group of five or more members above the redshift of $z>0.49$ is quite low, at about 5\%. In the case of KUV~00311-1938, the probability is somewhat higher, approximately 15\%, primarily due to the smaller size of the `groups' above $z>0.5$. Nonetheless, it is important to consider these as conservative estimates. As shown in Fig. \ref{fig:zmag_histogram}, our galaxy sample is not complete in the field, particularly at higher redshifts, which makes it likely that we are missing members from the group candidates.

\subsection{Prospects for using KUV~00311-1938 and S2~0109+22 to detect warm-hot intergalactic medium}

A significant portion of the baryons predicted by the current leading cosmological theory ($\lambda$CDM) remains undetected \citep[e.g.][]{shull2012,Danforth2016}. Cosmological simulations indicate that these missing baryons reside in the WHIM phase embedded in the filaments of the Cosmic Web \citep[e.g.][]{cen99,martizzi2019}. However, due to its low density, the expected X-ray emission from the WHIM is at or beyond the capabilities of current instrumentation. Instead, the WHIM could be detected as an absorption feature in the X-ray spectra of bright blazars located behind the WHIM filaments \citep[see e.g.][for possible detections]{fang2007, bonamente2016, ahoranta2020}.

WHIM filaments can be traced by galaxy groups in SDSS up to $z=0.155$, but the sample is only homogeneous up to $z=0.05-0.06$ \citep{tempel14}. Therefore, although our observations indicate the presence of foreground groups or structures at $z=0.15-0.26$ for both blazars, it is impossible to ascertain if these groups trace filamentary structures with sufficient WHIM column density. Significant improvement in this regard is expected with the 4MOST 4HS survey, where the chances of intercepting an absorbing system are anticipated to increase to a level of $\sim$50\% per sight line \citep{Tuominen2023}. 

Both blazars are X-ray bright with fluxes of F\,(0.3--10\,keV)$=7.2\times10^{-12}$ erg$/$cm$^2/$s \citep[KUV~00311-1938,][]{2010MNRAS.401.1570T} and F\,(0.3--10 keV)$= (2.5-15)\times10^{-12}$ erg$/$cm$^2/$s \citep[S2~0109+22,][]{2018MNRAS.480..879M}. These fluxes are sufficiently high to observe X-ray absorption features, similar to, for example, H~2356-309, which has been utilized in such studies by \citet{zappacosta10}. Consequently, both blazars could be intriguing sources for investigating WHIM at higher redshifts with new or upcoming X-ray missions such as XRISM and NewAthena.

\section{Conclusions}

In this paper, we identified several galaxy group candidates in the fields of both blazars, KUV~00311-1938 and S2~0109+22, through both our spectroscopic observations and our analysis of the distribution of photometric redshifts. We utilized previous direct optical spectroscopy limits to discern which of these groups are likely foreground groups. In both cases, we also observed galaxies with higher redshifts that could potentially be associated with the blazar.

Even before our study, KUV~00311-1938 was known to be within a redshift range where only a handful of VHE $\gamma$-ray emitting objects are available for indirect EBL studies. We did not detect any galaxies at redshift $z=0.506$, which is the lower limit from direct spectroscopy based on the Mg~II absorption line. Most of the galaxies we observed in spectroscopy turned out to be members of foreground groups. Intriguingly, our KDE analysis of the photometric redshifts suggests the presence of an additional galaxy group at $z\sim0.68$. Two of the galaxies we observed have spectroscopic redshifts $z=0.639-0.640$, but since they are only two, the probability of a chance coincidence is high ($\sim$15\%). A spectroscopic follow-up of the galaxies with photometric redshifts of $z\sim0.68$ could confirm this high redshift.

For S2~0109+22, we found a group of five galaxies with a redshift of $z=0.49$, and the chance probability for this association is low ($\sim$5\%). This association is further supported by our KDE analysis results, where we also identified this group. Both in spectroscopic and photometric analysis, the weighted central points of the candidate group lie very close to the blazar. Therefore, we conclude that the blazar very likely belongs to this candidate group and has a redshift of $z=0.49$.

In summary, we have demonstrated in this work that combining spectroscopic redshifts with an analysis of the photometric redshifts of galaxies around the blazar with an unknown redshift allowed us to determine the likely redshifts for both blazars. In the case of KUV~00311-1938, further confirmation of this high redshift could be achieved with additional observations. Many of the galaxies we used in this work have r-band magnitudes larger than 20, and therefore, spectroscopy still requires dedicated observations with rather large telescopes. Nevertheless, with upcoming surveys such as 4MOST coming online in the next few years, this method of determining the redshifts will become feasible for a larger sample of blazars. This will also help indentify the most suitable blazars for studies of extragalactic background light and the warm-hot intergalactic medium.

\section*{Acknowledgements}

Authors would like to thank Simona Paiano for providing the spectrum of S2~0109+22 and Francesco Massaro for discussions on applicability of his mock sample results to larger redshifts. We also thank Jukka Nevalainen for discussion on detecting the warm-hot intergalactic medium.
This project has received funding from the European Research Council (ERC) under the European Union’s Horizon 2020 research and innovation programme (grant agreement No. 101002352, PI: M. Linares). 
K.I.I.K. acknowledges the financial support from the visitor and mobility program of the Finnish Centre for Astronomy with ESO (FINCA), funded by the Academy of Finland grant nr 306531. 
E.L. was supported by Academy of Finland projects 317636, 320045, and 346071.
This research is based on observations collected at the European Southern Observatory under ESO programme 0100.B-0833.

\section*{Data Availability}

The VLT/FORS2 data analyzed here are available at
the European Southern Observatory Science Archive Facility
(\url{http://archive.eso.org}). The Gemini data is available at the Gemini Observatory archive (\url{https://archive.gemini.edu}). The SDSS data is available at the SDSS website: \url{https://www.sdss4.org}. The simulation data produced as part of this work are available from the authors on request.



\bibliographystyle{mnras}
\bibliography{redshift} 




\appendix

\section{Galaxy distance table}

\begin{table*}
    \caption{Properties of the galaxies in the field of KUV~00311-1938. Columns are (1) right ascension in degrees, (2) declination is degrees, (3) SDSS z'-band magnitude, (4) redshift from multi-object spectroscopy, (5) instrument used to obtain the spectrum, (6) SDSS photometric redshift, (7) separation to blazar in arcseconds, and (8) source numbering shown in Fig. \ref{fig:kuv_mosaic}.} 
    \centering
    \begin{tabular}{c|ccccccl}   
    \hline
    RA & DEC & z'$_{\rm AB}$ & SpecZ & Inst. & PhotoZ & $\Delta\theta$ & N \\
    (deg) & (deg) & (mag) & & & SDSS & ('') \\
    \hline
    8.348750 & -19.34886 & 21.05$\pm$0.17 & 0.4554 & OSIRIS & 0.49$\pm$0.17 & 156 & \#24  \\
    8.349691 & -19.33609 & 17.83$\pm$0.03 & 0.1480 & GMOS   & 0.12$\pm$0.03 & 170 & G\#2  \\
    8.354167 & -19.37567 & 21.94$\pm$0.12 & 0.2376 & OSIRIS & 0.25$\pm$0.12 & 145 & P\#23 \\
    8.357860 & -19.37570 & 20.07$\pm$0.08 & 0.4680 & GMOS   & 0.45$\pm$0.08 & 134 & G\#8  \\ 
    8.358805 & -19.39175 & 21.38$\pm$0.13 & 0.4126 & FORS2  & 0.32$\pm$0.13 & 166 & \#28  \\
    8.359580 & -19.36220 & 20.75$\pm$0.05 & 0.6400 & GMOS   & 0.67$\pm$0.05 & 115 & G\#7  \\
    8.360320 & -19.33116 & 16.70$\pm$0.01 & 0.1470 & GMOS   & 0.17$\pm$0.01 & 151 & G\#24 \\
    8.364630 & -19.39061 & 19.73$\pm$0.06 & 0.4140 & FORS2  & 0.44$\pm$0.06 & 149 & \#27  \\
    8.368750 & -19.39239 & 20.71$\pm$0.11 & 0.4120 & OSIRIS & 0.58$\pm$0.11 & 146 & P\#20 \\
    8.369520 & -19.35136 & 19.23$\pm$0.13 & 0.2900 & FORS2  & 0.35$\pm$0.13 & 85  & \#11  \\
    8.371250 & -19.34828 & 22.76$\pm$0.12 & 0.5350 & OSIRIS & 0.92$\pm$0.12 & 84  & P\#19 \\
    8.372214 & -19.39001 & 20.50$\pm$0.09 & 0.4830 & FORS2  & 0.47$\pm$0.09 & 132 & \#26  \\
    8.372987 & -19.38060 & 21.97$\pm$0.74 & 0.4235 & FORS2  &               & 103 & \#25  \\
    8.375833 & -19.36253 & 20.56$\pm$0.10 & 0.4134 & OSIRIS & 0.39$\pm$0.10 & 60  & P\#17 \\
    8.376025 & -19.33282 & 20.00$\pm$0.09 & 0.2190 & FORS2  & 0.40$\pm$0.10 & 112 & \#23  \\
    8.376279 & -19.35206 & 20.73$\pm$0.11 & 0.1605 & GMOS   & 0.39$\pm$0.11 & 63  & G\#13 \\
    8.379091 & -19.35232 & 18.18$\pm$0.03 & 0.1605 & FORS2  & 0.10$\pm$0.03 & 54  & \#7   \\
    8.383750 & -19.34083 & 21.93$\pm$0.21 & 0.1469 & OSIRIS & 0.47$\pm$0.21 & 74  & P\#16 \\
    8.384832 & -19.38943 & 20.60$\pm$0.07 & 0.1360 & FORS2  & 0.18$\pm$0.07 & 113 & \#24  \\
    8.386320 & -19.33076 & 20.96$\pm$0.09 & 0.3520 & GMOS   & 0.42$\pm$0.09 & 105 & G\#15 \\
    8.393923 & -19.35252 & 19.42$\pm$0.04 & 0.2373 & FORS2  & 0.20$\pm$0.04 & 24  & \#6   \\
    8.397430 & -19.32274 & 19.62$\pm$0.08 & 0.3855 & GMOS   & 0.46$\pm$0.08 & 132 & G\#18 \\
    8.402887 & -19.38667 & 19.59$\pm$0.13 & 0.1085 & FORS2  & 0.24$\pm$0.13 & 104 & \#9   \\
    8.408049 & -19.31351 & 19.41$\pm$0.04 & 0.3330 & GMOS   & 0.34$\pm$0.04 & 172 & G\#21 \\
    8.408534 & -19.37380 & 19.84$\pm$0.08 & 0.6390 & FORS2  & 0.58$\pm$0.08 & 74  & \#21  \\
    8.409522 & -19.39235 & 20.04$\pm$0.18 & 0.5520 & FORS2  & 0.37$\pm$0.18 & 132 & \#22  \\
    8.415103 & -19.36052 & 19.88$\pm$0.07 & 0.3260 & FORS2  & 0.44$\pm$0.07 & 74  & \#19  \\
    8.415633 & -19.33170 & 19.23$\pm$0.12 & 0.2400 & FORS2  & 0.37$\pm$0.12 & 125 & \#17  \\
    8.417479 & -19.37740 & 17.88$\pm$0.05 & 0.2180 & FORS2  & 0.16$\pm$0.05 & 105 & \#8   \\
    8.418285 & -19.34955 & 21.41$\pm$0.07 & 0.3850 & FORS2  & 0.24$\pm$0.07 & 92  & \#18  \\
    8.418388 & -19.31421 & 21.05$\pm$0.16 & 0.4760 & FORS2  & 0.49$\pm$0.16 & 183 & \#1   \\
    8.420417 & -19.34903 & 20.28$\pm$0.08 & 0.5800 & OSIRIS & 0.53$\pm$0.08 & 99  & P\#7  \\
    8.420970 & -19.31129 & 20.73$\pm$0.16 & 0.4750 & GMOS   & 0.45$\pm$0.16 & 197 & G\#40 \\
    8.422495 & -19.33668 & 18.10$\pm$0.03 & 0.1468 & FORS2  & 0.10$\pm$0.03 & 128 & \#3   \\
    8.428500 & -19.36273 & 19.40$\pm$0.10 & 0.2400 & GMOS   & 0.29$\pm$0.10 & 120 & G\#30 \\
    8.429510 & -19.33802 & 20.26$\pm$0.05 & 0.6550 & GMOS   & 0.65$\pm$0.05 & 145 & G\#28 \\
    8.432358 & -19.35128 & 17.57$\pm$0.02 & 0.1113 & FORS2  & 0.12$\pm$0.02 & 136 & \#4   \\
    8.434077 & -19.33986 & 20.65$\pm$0.12 & 0.5330 & FORS2  & 0.30$\pm$0.12 & 155 & \#15  \\
    8.450750 & -19.35922 & 20.44$\pm$0.12 & 0.4110 & FORS2  & 0.36$\pm$0.12 & 195 & \#14  \\    
    \end{tabular}
    \label{tab:kuv_table}
\end{table*}

\begin{table*}
    \caption{Properties of the galaxies in the field of S2~0109$+$22. Columns are (1) right ascension in degrees, (2) declination is degrees, (3) SDSS z'-band magnitude, (4) redshift from multi-object spectroscopy, (5) instrument used to obtain the spectrum, (6) SDSS photometric redshift, (7) separation to blazar in arcseconds, and (8) source numbering shown in Fig. \ref{fig:S2_mosaic}.} 
    \centering
    \begin{tabular}{c|ccccccl}   
    \hline
    RA & DEC & z'$_{\rm AB}$ & SpecZ & Inst. & PhotoZ & $\Delta\theta$ & N \\
    (deg) & (deg) & (mag) & & & SDSS & ('') \\
    \hline
    17.97224 & 22.71662 & 18.45$\pm$0.05 & 0.1835 & FORS2 & 0.14$\pm$0.03 & 199 & \#29  \\
    17.98279 & 22.79652 & 18.51$\pm$0.04 & 0.330  & FORS2 & 0.37$\pm$0.05 & 233 & \#13  \\
    17.98365 & 22.71332 & 18.02$\pm$0.02 & 0.1835 & FORS2 & 0.19$\pm$0.03 & 174 & \#17  \\
    17.98917 & 22.75067 & 20.20$\pm$0.15 & 0.4901 & GMOS  & 0.45$\pm$0.13 & 119 & G\#20 \\
    17.99043 & 22.78776 & 19.79$\pm$0.11 & 0.1840 & FORS2 & 0.28$\pm$0.13 & 193 & \#12  \\
    17.99250 & 22.78633 & 20.63$\pm$0.16 & 0.6630 & GMOS  & 0.54$\pm$0.10 & 185 & G\#29 \\
    17.99583 & 22.74864 & 19.78$\pm$0.10 & 0.7050 & GMOS  & 0.52$\pm$0.07 & 95  & G\#19 \\
    17.99606 & 22.69881 & 17.72$\pm$0.03 & 0.0887 & FORS2 & 0.10$\pm$0.04 & 188 & \#15  \\ 
    17.99691 & 22.76845 & 20.37$\pm$0.13 & 0.2665 & FORS2 & 0.25$\pm$0.07 & 126 & \#9  \\ 
    17.99833 & 22.76058 & 21.30$\pm$0.29 & 0.8570 & GMOS  & 0.88$\pm$0.22 & 104 & G\#27 \\
    18.00167 & 22.70736 & 19.87$\pm$0.12 & 0.3489 & GMOS  & 0.40$\pm$0.10 & 152 & G\#30 \\
    18.00417 & 22.70047 & 20.68$\pm$0.20 & 0.4567 & GMOS  & 0.35$\pm$0.13 & 171 & G\#31 \\
    18.00583 & 22.74936 & 20.58$\pm$0.12 & 0.6880 & GMOS  & 0.92$\pm$0.04 & 63  & G\#38 \\
    18.00958 & 22.78503 & 20.94$\pm$0.24 & 0.1315 & GMOS  & 0.30$\pm$0.08 & 155 & G\#23 \\
    18.01500 & 22.75117 & 21.86$\pm$0.39 & 0.6630 & GMOS  &               & 40  & G\#36 \\
    18.01625 & 22.75006 & 20.15$\pm$0.11 & 0.2673 & GMOS  & 0.38$\pm$0.11 & 34  & G\#9  \\
    18.01750 & 22.75922 & 20.46$\pm$0.13 & 0.3900 & GMOS  & 0.38$\pm$0.07 & 59  & G\#10 \\
    18.02375 & 22.75386 & 19.36$\pm$0.08 & 0.2694 & GMOS  & 0.36$\pm$0.07 & 35  & G\#6  \\
    18.02500 & 22.74794 & 17.43$\pm$0.02 & 0.2674 & GMOS  & 0.32$\pm$0.02 & 14  & G\#3  \\
    18.02625 & 22.74583 & 20.35$\pm$0.11 & 0.2685 & GMOS  & 0.24$\pm$0.13 & 9   & G\#5  \\
    18.02702 & 22.77615 & 18.25$\pm$0.05 & 0.1230 & FORS2 & 0.06$\pm$0.04 & 116 & \#10  \\ 
    18.02750 & 22.74100 & 20.33$\pm$0.14 & 0.4950 & GMOS  & 0.56$\pm$0.06 & 16  & G\#7  \\
    18.03042 & 22.74886 & 20.73$\pm$0.17 & 0.4915 & GMOS  &               & 27  & G\#35 \\
    18.03154 & 22.78269 & 17.06$\pm$0.02 & 0.1230 & FORS2 & 0.14$\pm$0.03 & 141 & \#11  \\ 
    18.03758 & 22.74753 & 18.13$\pm$0.03 & 0.2662 & FORS2 & 0.25$\pm$0.03 & 46  & \#21  \\
    18.04417 & 22.73317 & 22.17$\pm$0.43 & 0.3345 & GMOS  &               & 77  & G\#18 \\
    18.04583 & 22.75786 & 20.68$\pm$0.21 & 0.4950 & GMOS  & 0.48$\pm$0.10 & 87  & G\#24 \\
    18.04792 & 22.73986 & 19.88$\pm$0.10 & 0.4930 & GMOS  & 0.44$\pm$0.05 & 80  & G\#17 \\
    18.05326 & 22.77338 & 19.11$\pm$0.09 & 0.3500 & FORS2 & 0.41$\pm$0.09 & 143 & \#25  \\
    18.05339 & 22.74637 & 18.31$\pm$0.03 & 0.1950 & FORS2 & 0.24$\pm$0.04 & 97  & \#22  \\
    18.05458 & 22.75225 & 20.43$\pm$0.12 & 0.2685 & GMOS  &               & 105 & G\#12 \\
    18.05865 & 22.74699 & 21.82$\pm$0.42 & 0.1860 & GMOS  & 0.25$\pm$0.12 & 115 & G\#37 \\
    18.06000 & 22.78244 & 20.20$\pm$0.16 & 0.1867 & GMOS  & 0.23$\pm$0.10 & 182 & G\#21 \\
    18.06671 & 22.74903 & 18.27$\pm$0.04 & 0.1855 & FORS2 & 0.17$\pm$0.03 & 142 & \#23  \\
    18.06833 & 22.76264 & 20.70$\pm$0.28 & 0.5170 & GMOS  & 0.37$\pm$0.06 & 161 & G\#16 \\
    18.06942 & 22.73990 & 18.00$\pm$0.04 & 0.1865 & FORS2 & 0.18$\pm$0.04 & 151 & \#2   \\
    18.07092 & 22.76468 & 19.61$\pm$0.08 & 0.5155 & FORS2 & 0.30$\pm$0.09 & 172 & \#8   \\
    18.07358 & 22.77303 & 18.69$\pm$0.05 & 0.4100 & FORS2 & 0.38$\pm$0.03 & 194 & \#27  \\
    \end{tabular}
    \label{tab:s2_table}
\end{table*}


\bsp	
\label{lastpage}
\end{document}